\documentclass[journal]{IEEEtran}
\usepackage[latin1]{inputenc}
\usepackage{amsmath}
\usepackage{graphicx}
\usepackage{amssymb}
\usepackage{xcolor}
\makeatletter

\providecommand{\LyX}{L\kern-.1667em\lower.25em\hbox{Y}\kern-.125emX\@}

\def\BibTeX{{\rm B\kern-.05em{\sc i\kern-.025em b}\kern-.08em
    T\kern-.1667em\lower.7ex\hbox{E}\kern-.125emX}}

\makeatother
\begin{document}
\def\ZZ{{\mathbb Z}}
\def\RR{{\Bbb R}}
\def\NN{{\mathbb N}}
\def\CC{{\mathbb C}}

\title{On Time-Delay Compensators for Delayed-Output Systems}
	\author{Hieu Trinh
	\thanks{H.~Trinh is with the School of Engineering, Deakin University, Waurn Ponds, 75 Pigdons Road, Geelong, Australia. (email:  hieu.trinh@deakin.edu.au)}
}

\maketitle \maketitle \maketitle \thispagestyle{plain}
\pagestyle{plain}

\begin{abstract}
This paper advances the practical utility of functional observer theory by addressing sensing latency in linear time-delay systems. We address the estimation of the functional $z(t)=Fx(t)$ in cases where the measurement delay $h$ is independent of the internal state delay $\tau$, with a specific focus on the condition $0 < h < \tau$. To compensate for sensing lags, we propose a  functional observer structure characterized by multiple internal delays and an augmented architecture. Algebraic existence conditions are established alongside a constructive synthesis procedure. By incorporating an additional delayed measurement vector, we demonstrate that this approach significantly expands the design space and is applicable to a wider class of systems with larger state and output delays.

\end{abstract}


\section{System Description and Problem Statement}
We consider the following linear time-delay system with a constant time delay
\begin{align}
	\label{c2.1a}
	&\dot{x}(t)=Ax(t)+A_{\tau}x(t-\tau)+Bu(t),\\
	\label{c2.1b}
	&x(t)=\phi(t), \quad \forall t \in [-\tau, 0],
\end{align}
where $x(t)\in \mathbb{R}^n$ is the state vector,  $u(t)\in \mathbb{R}^r$ is the control input vector, and $\phi(t)$ is the initial function. The scalar $\tau>0$ is the time delay, $A, A_{\tau}\in \mathbb{R}^{n\times n}$ and $B\in \mathbb{R}^{n\times r}$ are constant matrices.

We consider a practical situation where the output measurement vector, $y(t)\in \mathbb{R}^{p}$, contains a linear function of only the \textit{delayed} state vector, $x(t-h)$, where $h>0$ is a constant time delay. Thus, the output measurement vector is defined as follows
\begin{align}
	\label{c2.2}
	y(t)=C_hx(t-h),
\end{align}
where $C_h\in \mathbb{R}^{p\times n}$, $0<p\leq n$, is a full-row rank matrix.

Specifically, this paper considers the case where the output measurement vector $y(t) = C_hx(t-h)$ is subject to a time-delay $h$ within the bounded interval $0 < h < \tau$. 

Note that the case where $h \ge \tau$ was recently investigated in \cite{trinhnam1}. Consequently, the results presented in this paper complement the findings reported there.

We are interested in designing an asymptotic observer to estimate the following functional
\begin{align}
	\label{c2.3}
	z(t)=Fx(t),
\end{align}
where $z(t)\in\mathbb{R}^m$ is a linear function of the instantaneous state vector, $x(t)$, and $F\in\mathbb{R}^{m\times n}$ is full row rank.

\textit{Remark 1:} For  $0 < h < \tau$, we can define a delayed output vector $y_{\alpha}(t)$ as follows
$$y_{\alpha}(t) = y(t - \alpha)$$where $\alpha = \tau - h > 0$. In this formulation, $y_{\alpha}(t)$ is simply the original output $y(t)$ delayed by $\alpha$. By utilizing $y_{\alpha}(t)$ to design a functional observer for estimating $z(t)$, we effectively synchronize the system; the delays present in both the state and output vectors become identical (see, \cite{trinhnam1}). Designing a functional observer under these conditions is more straightforward, as it reduces the complexity typically introduced by mismatched delay terms.

In contrast, this paper investigates the estimation of the functional (\ref{c2.3}) under mismatched delay conditions. Our objective is to construct an asymptotic observer, $\hat{z}(t)$, that ensures $\hat{z}(t) \to z(t)$ as $t \to \infty$. By explicitly distinguishing between these two delays, we demonstrate that it is possible to gain significant flexibility in satisfying the existence conditions for the functional observer.

Following the insights presented in Section V of reference \cite{trinhnam1},  we now utilize the following augmented measurement vector
\begin{align}
	\label{c2.4}
	y_a(t)&=\begin{pmatrix} y(t) \\ y_\alpha (t) \end{pmatrix}=\begin{pmatrix} y(t) \\ y(t-\alpha) \end{pmatrix}=\begin{pmatrix} C_hx(t-h) \\ C_h(t-\tau) \end{pmatrix}\nonumber\\&=\bar{C}_hx(t-h)+\bar{C}_{\tau}x(t-\tau),\end{align}
where $y_a(t)\in\mathbb{R}^{2p}$ and
\[\bar{C}_h:=\begin{pmatrix}
	C_h\\\mathbf 0\end{pmatrix}, \quad \bar{C}_{\tau}:=\begin{pmatrix}
	\mathbf 0\\ C_h\end{pmatrix}.\]

To estimate the functional (\ref{c2.3}) for system (\ref{c2.1a})-(\ref{c2.1b}) utilizing the delayed output vector $y_a(t)$ as defined by (\ref{c2.4}), we consider the following observer with multiple delayed terms
\begin{align}
	\hat{z}(t)&=w(t)+My_a(t)+M_{\tau}y_a(t-\tau),\label{c2.5a}\\
	\dot{w}(t)&=Nw(t)+N_hw(t-h)+N_{\tau}w(t-\tau)+Gy_a(t)\nonumber\\&+G_hy_a(t-h)+G_{\tau}y_a(t-\tau)+G_{\tau h}y_a(t-\tau -h)\nonumber\\&+G_{\tau \tau}y_a(t-2\tau)+Ju(t)+J_hu(t-h)+J_{\tau}u(t-\tau)\nonumber\\&+J_{\tau h}u(t-\tau-h)+J_{\tau \tau}u(t-2\tau),\label{c2.5b}
\end{align}
where $w(\theta)=\rho(\theta),\ \forall \theta \in[-\tau,0]$,  $w(t) \in \mathbb{R}^m$, and $\hat{z}(t)$ is the estimate of $z(t)$. Matrices $M$, $M_{\tau}$, $N$, $N_h$, $N_\tau$, $G$,  $G_h$, $G_\tau$,  $G_{\tau h}$, $G_{\tau \tau}$, $J$, $J_h$, $J_\tau$, $J_{\tau h}$ and $J_{\tau \tau}$ are to be determined such that $\hat{z}(t)$ converges asymptotically to $z(t)$. 

\textit{Remark 2:} The architecture of the observer described in equations (\ref{c2.5a})-(\ref{c2.5b}) utilizes multiple delayed components to achieve estimation accuracy, specifically addressing the challenge of mismatched latencies within the state and output vectors. Moreover, by having both terms $N_hw(t-h)$ and $N_{\tau}w(t-\tau)$ (where $\tau > h$), the design relaxes the stabilizability conditions for the estimation-error system. As will be shown later, this approach allows for larger allowable delay margins for $\tau$ and $h$.

\section{Observer Existence Conditions}
We derive existence conditions of observer (\ref{c2.5a})-(\ref{c2.5b}). Defining the estimation error vector $e(t)=\hat z(t)-z(t)$, the error dynamics are given by
\begin{align}
	\label{c2.6}
	\dot{e}(t)&=\dot{w}(t)+M\dot{y}_a(t)+M_{\tau}\dot{y}_a(t-\tau)-F\dot{x}(t)\nonumber\\  &=Ne(t)+N_he(t-h)+N_{\tau}e(t-\tau)+\mathcal{C}_{1}u(t)\nonumber\\  &+\mathcal{C}_{2}u(t-\tau) +\mathcal{C}_{3}u(t-h)+\mathcal{C}_{4}u(t-\tau -h) \nonumber\\  &+\mathcal{C}_{5}u(t-2\tau)+\mathcal{C}_{6}x(t)+\mathcal{C}_{7}x(t-h)+\mathcal{C}_{8}x(t-\tau)\nonumber\\ &+\mathcal{C}_{9}x(t-\tau-h)+\mathcal{C}_{10}x(t-2\tau)+\mathcal{C}_{11}x(t-2\tau-h)\nonumber\\  &+\mathcal{C}_{12}x(t-3\tau)
	+\mathcal{C}_{13}x(t-2h)+\mathcal{C}_{14}x(t-\tau-2h),
\end{align}
where\\ 
\\
$\mathcal{C}_1 =J-FB,\ \mathcal{C}_2 = J_{\tau}+M\bar{C}_{\tau}B, \ \mathcal{C}_3 = J_h+M\bar{C}_hB, \ \mathcal{C}_4 = J_{\tau h}+M_{\tau}\bar{C}_hB, \ \mathcal{C}_5 = J_{\tau \tau}+M_{\tau}\bar{C}_{\tau}B, \ \mathcal{C}_6 =NF-FA$, \
$\mathcal{C}_7 = N_hF+\bar{G}\bar{C}_h+M\bar{C}_hA$, \ $\mathcal{C}_8 = N_{\tau}F+\bar{G}\bar{C}_{\tau}+M\bar{C}_{\tau}A-FA_{\tau}$, \ $\mathcal{C}_9= \hat{G}_{\tau}\bar{C}_h+\bar{G}_h\bar{C}_{\tau}+M\bar{C}_hA_{\tau}+M_{\tau}\bar{C}_hA$, \ $\mathcal{C}_{10} = \hat{G}_{\tau}\bar{C}_{\tau}+M\bar{C}_{\tau}A_{\tau}+M_{\tau}\bar{C}_{\tau}A$, \ $\mathcal{C}_{11} = \bar{G}_{\tau \tau}\bar{C}_h+\bar{G}_{\tau h}\bar{C}_{\tau}+M_{\tau}\bar{C}_hA_{\tau},$ \ $\mathcal{C}_{12} = \bar{G}_{\tau \tau}\bar{C}_{\tau}+M_{\tau}\bar{C}_{\tau}A_{\tau},$ \ $\mathcal{C}_{13} = \bar{G}_h\bar{C}_h$, \ $\mathcal{C}_{14} = \bar{G}_{\tau h}\bar{C}_h$, \\
\\
$\bar{G}:=G-NM$, \ $\hat{G}_{\tau}:=G_{\tau}-N_{\tau}M-NM_{\tau}$, \ $\bar{G}_h:=G_h-N_hM$, \ $\bar{G}_{\tau \tau}:=G_{\tau \tau}-N_{\tau}M_{\tau}$, \ $\bar{G}_{\tau h}:=G_{\tau h}-N_hM_{\tau}$ \ and \
$\mathcal C
:=
\begin{pmatrix}
	\mathcal C_1 &
	\mathcal C_2 &
	& \cdots &
	\mathcal C_{13} &
	\mathcal C_{14}
\end{pmatrix}.
$

The following theorem provides conditions for the existence of observer (\ref{c2.5a})-(\ref{c2.5b}).

\textit{\textbf{Theorem 1:}}
	Observer (\ref{c2.5a})-(\ref{c2.5b}) provides asymptotic estimation of the functional $z(t)=Fx(t)$ and yields
	estimation error dynamics that are decoupled from the plant state $x(\cdot)$ and the input $u(\cdot)$
	if $\mathcal C=\bf 0$ and the following delay-dependent error dynamics
	\begin{align}
		\label{c2.7}&\dot{e}(t)=Ne(t)+N_he(t-h)+N_{\tau}e(t-\tau)
	\end{align}
	is asymptotically stable. In this case, the estimation error satisfies
	\[
	e(t)=\hat z(t)-z(t)\to {\bf 0} \quad \text{as} \quad t\to\infty
	\]
	for all admissible initial conditions and inputs $u(\cdot)$.

\begin{proof}	If $\mathcal C=\bf 0$, then \eqref{c2.6} reduces to \eqref{c2.7}, so the error dynamics are
	decoupled from $x(\cdot)$ and $u(\cdot)$. If, in addition, \eqref{c2.7} is asymptotically stable, then $e(t)\to \bf 0$ as $t\to\infty$ for all admissible initial conditions and inputs $u(\cdot)$. This completes the proof.
\end{proof}

\section{Determination of the Observer Parameters}
In this section, we derive observer parameters $M$, $M_{\tau}$, $N$, $N_h$, $N_\tau$, $G$, $G_h$, $G_\tau$, $G_{\tau h}$, $G_{\tau \tau}$, $J$,  $J_h$, $J_\tau$, $J_{\tau h}$ and $J_{\tau \tau}$ to satisfy the conditions presented in Theorem 1.

Let us first consider $\mathcal C=\bf 0$. We have
\[\mathcal C_1=\mathcal C_2=\mathcal C_3=\mathcal C_4=\mathcal C_5=\bf 0\] by the following direct
choices of $J$, $J_\tau$, $J_h$, $J_{\tau h}$ and $J_{\tau \tau}$, respectively,
\begin{align}
	\label{c2.8}
	&J = FB, \ J_{\tau}=-M\bar{C}_{\tau}B, \ J_h=-M\bar{C}_hB, \\ &J_{\tau h}=-M_{\tau}\bar{C}_hB, \  J_{\tau \tau}=-M_{\tau}\bar{C}_{\tau}B.
\end{align}

$\mathcal C_6=\bf 0$ iff
\begin{equation}
	\label{c2.9}
	\mathrm{rank}\begin{pmatrix}FA\\ F\end{pmatrix}=\mathrm{rank}(F).
\end{equation}
Subject to the satisfaction of condition (\ref{c2.9}), we obtain
\begin{equation}
	\label{c2.10}
	N=FAF^-.
\end{equation}
Let us next look at the requirement $\mathcal C_{13}=\bf 0$ and $\mathcal C_{14}=\bf 0$. For this, let $\bar{G}_h$ and $\bar{G}_{\tau h}$ be partitioned as follows
\[\bar{G}_h=\begin{pmatrix}\bar{G}_{h1} &\bar{G}_{h2}\end{pmatrix}, \quad \bar{G}_{\tau h}=\begin{pmatrix}\bar{G}_{\tau h1} &\bar{G}_{\tau h2}\end{pmatrix},\]
where $\bar{G}_{h1}$, $\bar{G}_{h2}$, $\bar{G}_{\tau h1}$ and $\bar{G}_{\tau h2}$ are of dimension $(m\times p)$.  
Given that $\bar{C}_h=\begin{pmatrix}
	C_h\\\mathbf 0\end{pmatrix}$, we have
\begin{align}
	\mathcal C_{13}&=\begin{pmatrix}\bar{G}_{h1} &\bar{G}_{h2}\end{pmatrix}\begin{pmatrix}
		C_h\\\mathbf 0\end{pmatrix},\nonumber \\ \mathcal C_{14}&=\begin{pmatrix}\bar{G}_{\tau h1} &\bar{G}_{\tau h2}\end{pmatrix}\begin{pmatrix}
		C_h\\\mathbf 0\end{pmatrix}.\nonumber
\end{align}
Since $C_h$ is full row rank, $\mathcal C_{13}=\mathbf 0$ and $\mathcal C_{14}=\mathbf 0$ iff $\bar{G}_{h1}=\bar{G}_{\tau h1}=\mathbf 0$. Thus, we obtain
\[\bar{G}_h=\begin{pmatrix}\mathbf 0 &\bar{G}_{h2}\end{pmatrix}, \quad \bar{G}_{\tau h}=\begin{pmatrix}\mathbf 0 &\bar{G}_{\tau h2}\end{pmatrix}.\]
where $\bar{G}_{h2}$ and $\bar{G}_{\tau h2}$ are arbitrary. Since $\bar{C}_{\tau}=\begin{pmatrix}
	\mathbf 0\\ C_h\end{pmatrix}$, we have 
\begin{align}
	&\bar{G}_h\bar{C}_{\tau}=\begin{pmatrix}\mathbf 0 &\bar{G}_{h2}\end{pmatrix}\begin{pmatrix}
		\mathbf 0\\ C_h\end{pmatrix}=\bar{G}_{h2}C_h,\nonumber\\
	\label{c2.10b}
	&\bar{G}_{\tau h}\bar{C}_{\tau}=\begin{pmatrix}\mathbf 0 &\bar{G}_{\tau h2}\end{pmatrix}\begin{pmatrix}
		\mathbf 0\\ C_h\end{pmatrix}=\bar{G}_{\tau h2}C_h.\nonumber
\end{align}
Substituting $\bar{G}_h\bar{C}_{\tau}$ with $\bar{G}_{h2}C_h$, the expression for $\mathcal C_{9}$ becomes
\[ \mathcal{C}_9= \hat{G}_{\tau}\bar{C}_h+\bar{G}_{h2}C_h+M\bar{C}_hA_{\tau}+M_{\tau}\bar{C}_hA.\]
Given the structure $\bar{C}_h = \begin{pmatrix} C_h \\ \mathbf{0} \end{pmatrix}$, it follows that the term $\bar{G}_{h2}C_h$ introduces no additional degrees of freedom to the constraint $\mathcal{C}_9 = \mathbf{0}$. A similar observation holds for $\mathcal{C}_{11}$ when $\bar{G}_{\tau h}\bar{C}_{\tau}$ is substituted with $\bar{G}_{\tau h2}C_h$. Therefore, we let 
\[\bar{G}_{h2}=\bar{G}_{\tau h2}=\mathbf 0.\]
With the above choice, we have $\bar{G}_h=\bar{G}_{\tau h}=\mathbf 0$, and hence $G_h$ and $G_{\tau h}$ are obtained as follows
\begin{align}
	\label{c2.11}
	G_h = N_hM, \quad G_{\tau h}=N_hM_{\tau}.
\end{align}

Next, we consider the requirement $\mathcal C_{7}=\mathcal C_{8}=\mathcal C_{9}=\mathcal C_{10}=\mathcal C_{11}=\mathcal C_{12}=\bf 0$, which can be written as the following linear matrix equation
\begin{align}
	\label{c2.12a}
	X\Theta=\Upsilon,
\end{align}
where $X\in\mathbb{R}^{m\times (2m+10p)}$, $\Theta\in\mathbb{R}^{(2m+10p)\times 6n}$ and $\Upsilon\in\mathbb{R}^{m\times 6n}$ are defined as follows
\begin{align}
	\label{c2.12b}
	X&=\begin{pmatrix} N_h &N_{\tau} &\bar{G} &\hat{G}_{\tau} &\bar{G}_{\tau \tau} &M &M_{\tau}
	\end{pmatrix},\\
	\label{c2.12c}
	\Theta &= \begin{pmatrix} F &\mathbf{0} &\mathbf{0} &\mathbf{0} &\mathbf{0} &\mathbf{0}\\\mathbf{0} &F &\mathbf{0} &\mathbf{0} &\mathbf{0} &\mathbf{0}\\\bar{C}_h &\bar{C}_{\tau} &\mathbf{0} &\mathbf{0} &\mathbf{0} &\mathbf{0}\\\mathbf{0} &\mathbf{0} &\bar{C}_h &\bar{C}_{\tau} &\mathbf{0} &\mathbf{0}\\\mathbf{0} &\mathbf{0} &\mathbf{0} &\mathbf{0} &\bar{C}_h &\bar{C}_{\tau}\\
		\bar{C}_hA &\bar{C}_{\tau}A &\bar{C}_hA_{\tau} &\bar{C}_{\tau}A_{\tau} &\mathbf{0} &\mathbf{0}\\\mathbf{0} &\mathbf{0} &\bar{C}_hA &\bar{C}_{\tau}A &\bar{C}_hA_{\tau} &\bar{C}_{\tau}A_{\tau}
	\end{pmatrix},\\
	\label{c2.12d}
	\Upsilon&=\begin{pmatrix} \mathbf{0} &FA_{\tau} &\mathbf{0} &\mathbf{0} &\mathbf{0} &\mathbf{0}
	\end{pmatrix}.
\end{align}
By Lemma 1 \cite{trinhnam1}, the matrix equation \eqref{c2.12a} is solvable 
if and only if
\begin{align}
	\label{c2.12e}
	\mathrm{rank}\begin{pmatrix}\Upsilon\\ \Theta\end{pmatrix}
	=\mathrm{rank}(\Theta).
\end{align}
Whenever this condition holds, the general solution of \eqref{c2.12a} according to Lemma 1 \cite{trinhnam1} is
\begin{equation}\label{c2.13}
	X=\Upsilon\Theta^{-}
	+Z\big(I_{2m+10p}-\Theta \Theta^{-}\big),
\end{equation}
where $Z\in\mathbb{R}^{m\times (2m+10p)}$ is arbitrary.

Partitioning \(X\) yields
\begin{align}\label{c2.13a}
	N_h=N_{h1}+ZN_{h2},
	\quad
	N_\tau=N_{\tau_1}+ZN_{\tau_2},
\end{align}
where
\begin{align}\label{c2.13b}
	N_{h1}&
	=
	\Upsilon\Theta^{-}\mathcal I_1,
	\quad
	N_{h2}
	=
	\big(I_{2m+10p}-\Theta \Theta^{-}\big)\mathcal I_1,\\
	\label{c2.13c}
	N_{\tau_1}&
	=
	\Upsilon\Theta^{-}\mathcal I_2,
	\quad
	N_{\tau_2}
	=
	\big(I_{2m+10p}-\Theta \Theta^{-}\big)\mathcal I_2,
\end{align}
with
\[
\mathcal I_1=
\begin{pmatrix}
	I_m\\
	\mathbf 0_{(m+10p)\times m}
\end{pmatrix},
\quad
\mathcal I_2=
\begin{pmatrix}
	\mathbf 0_{m\times m}\\
	I_m\\
	\mathbf 0_{10p\times m}
\end{pmatrix}.
\]
Hence the decoupled estimation error dynamics (\ref{c2.7}) takes the form
\begin{align}
	\label{c2.14}
	\dot e(t)
	&=Ne(t)
	+(N_{h1}+ZN_{h2})e(t-h)\nonumber\\&+(N_{\tau_1}+ZN_{\tau_2})e(t-\tau).
\end{align}

Therefore, if there exists a matrix \(Z\) such that the above
time-delay system is asymptotically stable, then observer (\ref{c2.5a})-(\ref{c2.5b}) provides asymptotic estimation of the functional $z(t)=Fx(t)$. Moreover, by Lemma 13 \cite{trinhnam2}, a sufficient condition for asymptotic
stability of this error time-delay system is the feasibility of the LMI in
Lemma 13 \cite{trinhnam2}.

Once asymptotic stability of (\ref{c2.14}) is ensured, $N_h$, $N_{\tau}$, $\bar G$, $\hat{G}_{\tau}$, $\bar{G}_{\tau \tau}$, $M$ and $M_{\tau}$
are then obtained directly from the corresponding blocks of $X$, and thus all the observer gains are obtained. 

\textit{Remark 3:} We discuss the case where for given $h$ and $\tau$, $\tau >h$, we first carry out the stabilization problem of finding $N_{h}$ and $N_{\tau}$ such that the following error time-delay system
\[\dot{e}(t)=Ne(t)+N_he(t-h)+N_{\tau}e(t-\tau),\]
$N=FAF^-$, is asymptotically stable. Here, as reported in \cite{trinhnam2}, a sufficient condition for asymptotic
stability of the above time-delay system is the feasibility of the LMI in Lemma
13 \cite{trinhnam2}.

Consequently, to satisfy the requirement that $\mathcal{C}_{i} = \mathbf{0}$ for $i \in \{7, \dots, 12\}$, and given the previously determined values for $N_h$ and $N_{\tau}$, the system can be expressed as the following linear matrix equation
\begin{align}
	\label{c2.15a}
	\bar{X}\bar{\Theta}=\bar{\Upsilon},
\end{align}
where
\begin{align}
	\label{c2.15b}\bar{X}&:=\begin{pmatrix}\bar{G} &\hat{G}_{\tau} &\bar{G}_{\tau \tau} &M &M_{\tau}
	\end{pmatrix},\\
	\label{c2.15c}
	\bar{\Theta}&:= \begin{pmatrix} \bar{C}_h &\bar{C}_{\tau} &\mathbf{0} &\mathbf{0} &\mathbf{0} &\mathbf{0}\\\mathbf{0} &\mathbf{0} &\bar{C}_h &\bar{C}_{\tau} &\mathbf{0} &\mathbf{0}\\\mathbf{0} &\mathbf{0} &\mathbf{0} &\mathbf{0} &\bar{C}_h &\bar{C}_{\tau}\\
		\bar{C}_hA &\bar{C}_{\tau}A &\bar{C}_hA_{\tau} &\bar{C}_{\tau}A_{\tau} &\mathbf{0} &\mathbf{0}\\\mathbf{0} &\mathbf{0} &\bar{C}_hA &\bar{C}_{\tau}A &\bar{C}_hA_{\tau} &\bar{C}_{\tau}A_{\tau}
	\end{pmatrix},\\
	\label{c2.15d}
	\bar{\Upsilon}&:=\begin{pmatrix} -N_hF &-N_{\tau}F+FA_{\tau} &\mathbf{0} &\mathbf{0} &\mathbf{0} &\mathbf{0}
	\end{pmatrix}.
\end{align}
By Lemma 1 \cite{trinhnam1}, the matrix equation \eqref{c2.15a} is solvable 
if and only if
\begin{align}
	\label{c2.15e}
	\mathrm{rank}\begin{pmatrix}\bar{\Upsilon}\\ \bar{\Theta}\end{pmatrix}
	=\mathrm{rank}(\bar{\Theta}).
\end{align}
Conversely, if $\bar{\Theta}$ has full column rank, i.e.,  $\text{rank}(\bar{\Theta}) = 6n$, the above required rank condition is always satisfied. Because $\bar{\Theta}$ is a known matrix, verifying this property is a straightforward task. If this condition is met, it allows us to prioritize the stabilization problem of the error time-delay system (\ref{c2.7}). As will be shown in the numerical example, addressing the stabilization of the error system first results in enhanced robust stability and a significantly larger allowable delay margins for $h$ and $\tau$.

A solution for $\bar{X}$ is as follows 
\begin{align}
	\label{c2.16}
	\bar{X}=\bar{\Upsilon}\bar{\Theta}^-,
\end{align}
and thus the rest of the observer gains are obtained.

\subsection{Synchronization of delays in the state and output vectors}
\label{subsec3}
As outlined in Remark 1, the delays in the state and output vectors can be synchronized by delaying $y(t)$ by $\alpha$ (where $\alpha = \tau - h > 0$). Utilizing this adjusted output, $y_{\alpha}(t)=y(t-\alpha)=C_hx(t-\tau)$, simplifies the functional observer design for $z(t)$ and eliminates the difficulties associated with mismatched delays \cite{trinhnam1}. Using $y_{\alpha}(t)$, let us now consider the following functional observer 
\begin{align}
	\label{c2.17a}\hat{z}(t)&=w(t)+My_{\alpha}(t),\\
	\dot{w}(t)&=Nw(t)+N_{\tau}w(t-\tau)+Gy_{\alpha}(t)+G_{\tau}y_{\alpha}(t-\tau)\nonumber\\&+Ju(t)+J_{\tau}u(t-\tau),\label{c2.17b}
\end{align}
where $w(\theta)=\rho(\theta),\ \forall \theta \in[-\tau,0]$,  $w(t) \in \mathbb{R}^m$, and $\hat{z}(t)$ is the estimate of $z(t)$. Matrices $M$, $N$, $N_{\tau}$, $G$, $G_{\tau}$, $J$ and $J_{\tau}$ are to be determined such that $\hat{z}(t)$ converges asymptotically to $z(t)$.

We now derive existence conditions of observer (\ref{c2.17a})-(\ref{c2.17b}). Defining the estimation error vector $e(t)=\hat z(t)-z(t)$, the error dynamics are given by
\begin{align}
	\label{c2.18}
	\dot{e}(t)&=\dot{w}(t)+M\dot{y}_{\alpha}(t)-F\dot{x}(t)\nonumber\\ \quad &=Ne(t)+N_{\tau}e(t-\tau)+\mathcal{\bar{C}}_{1}u(t)+\mathcal{\bar{C}}_{2}u(t-\tau) +\mathcal{\bar{C}}_{3}x(t)\nonumber\\ &+\mathcal{\bar{C}}_{4}x(t-\tau)+\mathcal{\bar{C}}_{5}x(t-2\tau),
\end{align}
where\\ 
\\
$\mathcal{\bar{C}}_1 =J-FB,\quad \mathcal{\bar{C}}_2 = J_{\tau}+MC_hB, \quad \mathcal{\bar{C}}_3 =NF-FA$, \quad
$\mathcal{\bar{C}}_4 = N_{\tau}F+\bar{G}C_h+MC_hA-FA_{\tau}$, \quad $\mathcal{\bar{C}}_5 =\bar{G}_{\tau}C_h+MC_hA_{\tau}$, \quad
$\bar{G}:=G-NM$, \quad $\bar{G}_{\tau}:=G_{\tau}-N_{\tau}M$, \ and\\
$\mathcal {\bar{C}}
:=
\begin{pmatrix}
	\mathcal {\bar{C}}_1 &
	\mathcal {\bar{C}}_2 &
	\mathcal {\bar{C}}_3 &
	\mathcal {\bar{C}}_4 &
	\mathcal {\bar{C}}_5
\end{pmatrix}.
$

\textit{\textbf{Theorem 2:}}
	Observer (\ref{c2.17a})-(\ref{c2.17b}) achieves asymptotic estimation of
	$z(t)=Fx(t)$,
	with estimation-error dynamics decoupled from the plant state and input vectors,
	if the following conditions hold.
	
	\begin{itemize}
		
		\item[(i)]
		The rank conditions
		\[
		\mathrm{rank}\begin{pmatrix}FA\\ F\end{pmatrix}=\mathrm{rank}(F),
		\]
		\[
		\mathrm{rank}\begin{pmatrix}\tilde{\Upsilon}\\ \tilde{\Theta}\end{pmatrix}
		=\mathrm{rank}(\tilde{\Theta}), 
		\]
		\[\quad \tilde{\Theta}:=\begin{pmatrix}F &\mathbf{0}\\ C_h &\mathbf{0}\\\mathbf{0} & C_h\\C_hA &C_hA_{\tau}\end{pmatrix}, \quad \tilde{\Upsilon}:=\begin{pmatrix}FA_{\tau} &\mathbf{0}\end{pmatrix},\]
		are satisfied.
		
		\item[(ii)]
		There exists a matrix \(Z\) such that the following delay-dependent error dynamics 	is asymptotically stable
		\[
		\dot e(t)
		= FAF^-e(t)
		+
		(N_{\tau_1}+Z N_{\tau_2})e(t-\tau),
		\]
		where
		\[
		N_{\tau_1}
		:=
		\tilde{\Upsilon}\tilde{\Theta}^{-}\mathcal I_1,
		\quad
		N_{\tau_2}
		:=
		\big(I_{m+3p}-\tilde{\Theta} \tilde{\Theta}^{-}\big)\mathcal I_1,
		\]
		
		and $\mathcal I_1$ denotes the first $m$ columns of $I_{m+3p}$.
		
	\end{itemize}
	In this case the estimation error satisfies
	\[
	e(t)\to \mathbf 0
	\qquad \text{as } t\to\infty
	\]
	for all admissible initial conditions and inputs \(u(\cdot)\).

\begin{proof}
	Under the decoupling condition $\bar{\mathcal C}=\mathbf 0$, the
	error equation \eqref{c2.18} reduces to
	\begin{align}
		\label{c2.19}
		\dot{e}(t)=Ne(t)+N_{\tau}e(t-\tau),\end{align}
	so that the estimation error dynamics are decoupled from the plant
	state $x(\cdot)$ and the input $u(\cdot)$.
	
	The condition $\mathcal {\bar{C}}_3=\bf0$ is equivalent to
	\begin{equation}\label{c2.20}
		NF=FA.
	\end{equation}
	By Lemma 1 \cite{trinhnam1}, \eqref{c2.20} admits a solution $N$ if and only if
	$\mathrm{rank}\begin{pmatrix}FA\\ F\end{pmatrix}=\mathrm{rank}(F)$, which gives the first condition in item~(i).
	Moreover, since $F$ is full row rank there exists a right inverse $F^{-}$ with $FF^{-}=I_m$,
	and any solution of \eqref{c2.20} satisfies $N=FAF^{-}$.
	
	The constraints $\mathcal {\bar{C}}_4=\bf0$ and $\mathcal {\bar{C}}_5=\bf0$ can be written as the linear matrix equation
	\[
	\tilde{X}\tilde{\Theta}=\tilde{\Upsilon},
	\]
	where $\tilde{X}:=\begin{pmatrix}N_{\tau} &\bar{G} &\bar{G}_{\tau} &M
	\end{pmatrix}$. The above equation is solvable if and only if
	\[
	\mathrm{rank}\begin{pmatrix}\tilde{\Upsilon}\\ \tilde{\Theta}\end{pmatrix}
	=\mathrm{rank}(\tilde{\Theta}),
	\]
	which establishes second condition in item~(i).
	
	Whenever this condition holds, the general solution yields
	\[N_\tau=N_{\tau_1}+Z N_{\tau_2},
	\]
	and hence the error dynamics takes the form
\begin{equation}\label{c2.20a}
	\dot e(t)
	= FAF^-e(t)
	+
	(N_{\tau_1}+Z N_{\tau_2})e(t-\tau).
\end{equation}
	By Lemma 11 \cite{trinhnam2}, a sufficient condition for asymptotic stability
	of this delay system is the feasibility of the LMI in Lemma 11 \cite{trinhnam2}.
	This establishes item~(ii), which completes the proof.
\end{proof}

The remaining observer matrices are then recovered from the
corresponding blocks of $\tilde{X}$, with the observer gains given by
\[
G=\bar G+NM, \ G_{\tau}=\bar G_{\tau}+N_{\tau}M, \
J=FB,\ J_{\tau}=-MC_hB.\]

\textit{Remark 4:} For the case where $\mathrm{rank}\begin{pmatrix}C_h &\mathbf{0}\\\mathbf{0} &C_h\\ C_hA &C_hA_{\tau}\end{pmatrix}=2n$, the constraints $\mathcal {\bar{C}}_4=\bf0$ and $\mathcal {\bar{C}}_5=\bf0$ can be written as the following linear matrix equation
\begin{align}
	\label{c2.21}
	X_r\Theta_r= \Upsilon_r,
\end{align}
where

$X_r:=\begin{pmatrix}\bar{G}  &\bar{G}_{\tau} &M\end{pmatrix}$, $ \Theta_r:=\begin{pmatrix}C_h &\mathbf{0}\\\mathbf{0} &C_h\\ C_hA &C_hA_{\tau}\end{pmatrix}$,

$\Upsilon_r=\begin{pmatrix}FA_{\tau}-N_{\tau}F   &\mathbf{0}\end{pmatrix}$.

Since $\mathrm{rank}(\Theta_r)=2n$, by Lemma 1 \cite{trinhnam1}, a solution
$X_r$ always exists for any $N_{\tau}\neq \bf 0$. In this case, Lemma 9 \cite{trinhnam2} can be used to stabilize (\ref{c2.20a}).

\section{Numerical Example}
We now present a numerical example to illustrate the proposed existence conditions and observer synthesis procedures.

\textit{Example:}
Consider the time-delay system with system matrices
\[
A=
\begin{pmatrix}
	-2 & 1\\
	0 & 1
\end{pmatrix},\quad
A_\tau=
\begin{pmatrix}
	-4 & 1\\
	2 & 1
\end{pmatrix},\quad
B=
\begin{pmatrix}
	1\\
	2
\end{pmatrix}.
\]
The delayed output measurement is given by
\[
y(t)=C_hx(t-h),
\]
where $C_h=I_2$.

We consider the following three cases to illustrate the results of this paper.

\textit{Case 1:} $\tau=1\text{s}$. It can be verified using the approach in
\cite{wu} that the time-delay system has unstable eigenvalues
located at $\{0.2802\pm j2.3965,  1.3907\}$.

Let $h=0.5\text{s}$, i.e., $h<\tau$, the delay affecting the output measurements is less than the delay associated with the state evolution.

Our objective is to estimate the
functional
\[
z(t)=Fx(t)=
\begin{pmatrix}
	0 & 1
\end{pmatrix}x(t)=x_2(t).
\]
In this case, we highlight the advantages of the design approach using observer (\ref{c2.5a})-(\ref{c2.5b}) over observer (\ref{c2.17a})-(\ref{c2.17b}).   

First, let us design an observer (\ref{c2.17a})-(\ref{c2.17b}) to estimate $z(t)$ using $y_{\alpha}(t)$. In this regard, as discussed in Section \ref{subsec3}, the delays in the state and output vectors are synchronized by delaying $y(t)$ by $0.5\text{s}$ so that $y_{\alpha}(t)=y(t-0.5)=x(t-1)$. With 
\[
F=
\begin{pmatrix}
	0 & 1
\end{pmatrix},
\]
$C_h=I_2$, and $A$ and $A_{\tau}$ given above, the rank conditions of Theorem 2 are satisfied. We obtain $N=FAF^{-}=1$, which is not Hurtwiz.

Furthermore, for this example, $\mathrm{rank}(\Theta_r)=4$, and thus we can first find $N_{\tau}$
to stabilize the following error time-delay system
\[
\dot e(t)
=e(t)+N_{\tau}e(t-\tau).
\]
According to \cite{mori}, the exact stability conditions for this
system are
\[
1+N_{\tau}<0, \qquad N_{\tau}\geq -\frac{1}{\tau}.
\]
These conditions imply that the delay $\tau$ must satisfy
\[
0<\tau<1\mathrm{s}.
\]
However, when $\tau\geq 1\mathrm{s}$, the estimation-error system
cannot be guaranteed to be asymptotically stable using observer (\ref{c2.17a})-(\ref{c2.17b}). Therefore, we consider an observer (\ref{c2.5a})-(\ref{c2.5b}) to estimate $z(t)$ using the following
augmented measurement vector
\begin{align}
	y_a(t)&=\begin{pmatrix} x(t-0.5) \\ x(t-1) \end{pmatrix}\nonumber\\&=\bar{C}_hx(t-0.5)+\bar{C}_{\tau}x(t-1),\nonumber \end{align}
where $y_a(t)\in\mathbb{R}^4$ and
\[\bar{C}_h=\begin{pmatrix}
	1 &0\\0 &1\\0 &0\\0 &0\end{pmatrix}, \quad \bar{C}_{\tau}=\begin{pmatrix}
	0 &0\\0 &0\\1 &0\\0 &1\end{pmatrix}.\]
Here, $y_a(t)$ contains the original delayed measurement vector, $y(t)=x(t-0.5)$, and a further delayed measurement vector, where $x(t-0.5)$ is being delayed by $0.5\text{s}$ to give $x(t-1)$.

With $\bar{C}_h$, $\bar{C}_{\tau}$, $A$ and $A_{\tau}$ given above, we can easily check that $\bar{\Theta}$ is full column rank since $\text{rank}(\bar{\Theta}) = 12$.  Thus, as discussed in Remark 3, we first carry out the stabilization problem of finding $N_{h}$ and $N_{\tau}$ such that the following error time-delay system
\begin{align}
	\label{c2.22}
	\dot{e}(t)=e(t)+N_he(t-0.5)+N_{\tau}e(t-1),\end{align}
is asymptotically stable. With $h$ and $\tau$ given as $h=0.5\mathrm{s}$ and $\tau=1\mathrm{s}$, the LMI in
Lemma 13 \cite{trinhnam2} is feasible and we obtain
\[N_h=-1.0476, \quad N_{\tau}=-0.2685.\]

Finally, with $h=0.5\mathrm{s}$, $\tau=1\mathrm{s}$ and by substituting $N_h=-1.0476$ and $N_{\tau}=-0.2685$ into (\ref{c2.15d}), we obtain $\bar{\Upsilon}$. Then from (\ref{c2.16}), we obtain $\bar{X}$, and hence the following first-order observer to estimate instantaneously the state variable $x_2(t)$
\begin{align}
	&\hat{x}_2(t)=w(t)+\begin{pmatrix}0.1340 &0.2595 &-0.1299 &0.0946\end{pmatrix}y_a(t),\nonumber\\&+\begin{pmatrix}-0.0429 &-0.0930 &0.0831 &0.0762\end{pmatrix}y_a(t-1)\nonumber\\
	&\dot{w}(t)=w(t)-1.0476w(t-0.5)-0.2685w(t-1)\nonumber\\&+\begin{pmatrix}0.4021 &0.9135 &1.6103 &1.3984 \end{pmatrix}y_a(t)\nonumber\\&+\begin{pmatrix}-0.1404 &-0.2718  &  0.1361  & -0.0991 \end{pmatrix}y_a(t-0.5)\nonumber\\&+\begin{pmatrix}-0.1473  & -0.4203  & -0.4247 &  -0.0731 \end{pmatrix}y_a(t-1)\nonumber\\&+\begin{pmatrix}0.0449   & 0.0974  & -0.0870 &  -0.0799 \end{pmatrix}y_a(t-1.5)\nonumber\\&+\begin{pmatrix}0.0260  &  0.1608  &  0.1574 &  -0.1798 \end{pmatrix}y_a(t-2)+2u(t)\nonumber\\&-0.653u(t-0.5)-0.0592u(t-1)+0.2288u(t-1.5)\nonumber\\&-0.2355u(t-2).\nonumber
\end{align}

\textit{Remark 5:} As discussed in Remark 2, by having both terms $N_he(t-h)$ and $N_{\tau}e(t-\tau)$, $\tau > h$, the stabilizability condition for the estimation-error system is relaxed. This approach allows for larger allowable delay margins for $\tau$ and $h$. For example, with $h=1.1\mathrm{s}$ and for a larger $\tau$, say, $\tau=2.2\mathrm{s}$, the LMI in
Lemma 13 \cite{trinhnam2} is still feasible and we obtain $N_h=-1.2031$, $N_{\tau}=0.1918$,
and the following error time-delay system
\[\dot{e}(t)=e(t)-1.2031e(t-1.1)+0.1918e(t-2.2),\]
is asymptotically stable. Note that observer (\ref{c2.17a})-(\ref{c2.17b}) which uses the measurement vector $y_{\alpha}(t)$ can only deal with delays $0 < h \le \tau <1\mathrm{s}$. This is a clear demonstration of the innovation inherent in the design of observer (\ref{c2.5a})-(\ref{c2.5b}) which utilizes measurement vector $y_a(t)$.

\textit{Case 2:} With the same $h$ and $\tau$ as in Case 1 (i.e., $h=0.5\mathrm{s}$ and $\tau=1\mathrm{s}$). In this case, we utilize an augmented output vector that incorporates fewer delayed measurements compared to Case 1. By reducing the number of delayed measurements in the augmented vector, we aim to simplify the observer structure while still being able to maintain an asymptotic estimation of $z(t)$.

Let us now consider the following
augmented measurement vector
\begin{align}
	\bar{y}_a(t)&=\begin{pmatrix} x(t-0.5) \\ x_1(t-1) \end{pmatrix}\nonumber\\&=\bar{C}_hx(t-0.5)+\bar{C}_{\tau}x(t-1),\nonumber \end{align}
where $\bar{y}_a(t)\in\mathbb{R}^3$ and
\[\bar{C}_h=\begin{pmatrix}
	1 &0\\0 &1\\0 &0\end{pmatrix}, \quad \bar{C}_{\tau}=\begin{pmatrix}
	0 &0\\0 &0\\1 &0\end{pmatrix}.\]
Thus, $\bar{y}_a(t)$ now contains one less delayed measurement than $y_a(t)$ as in Case 1. Here, we take $x_1(t-0.5)$ and delay it by $0.5\text{s}$ to give $x_1(t-1)$.

Our objective is to design an observer to estimate instantaneously the state variable $x_2(t)$ using the delayed output measurement vector $\bar{y}_a(t)$.

Now, with the above $\bar{C}_h$ and $\bar{C}_{\tau}$, we find that $\bar{\Theta}$ is not full column rank since $\text{rank}(\bar{\Theta})=11$.  Therefore, we need to solve the stabilization problem of finding $Z$ to ensure asymptotic stability of (\ref{c2.14}). To do this, we first need condition (\ref{c2.12e}) to be satisfied.

It is found that condition (\ref{c2.12e}) is satisfied, and from (\ref{c2.13b})-(\ref{c2.13c}), we obtain
\[N_{h1}=0, \quad N_{h2} \neq \mathbf{0}_{17\times 1}, \quad N_{\tau 1}=1, \quad N_{\tau 2}=\mathbf{0}_{17\times 1}.  \]

Since $ N_{h2} \neq \mathbf{0}$ and its first entry is non-zero, which is $0.6156$, we can simply choose $Z$ as follows 
\[Z=\begin{pmatrix} z_1 & \mathbf{0}_{1\times 16}\end{pmatrix},\] and we solve the problem of finding a $z_1$ such that the following error time-delay is asymptotically stable
\begin{align}
	\label{c2.23}
	\dot{e}(t)=e(t)+0.6156z_1e(t-0.5)+e(t-1),\end{align}
is asymptotically stable.

Because $N_{\tau 2} = \mathbf{0}$, finding a $z_1$ that ensures the asymptotic stability of (\ref{c2.23}) is a more restrictive stabilization problem than finding $N_h$ and $N_{\tau}$ for the error time-delay system (\ref{c2.22}). This increased conservatism is expected, as Case 2 provides one fewer measurement than Case 1.

By Lemma 13 \cite{trinhnam2}, the LMI is feasible and we obtain
\[z_1=-3.6108, \quad N_h=0.6156z_1=-2.2229.\]

Substituting $Z=\begin{pmatrix} -3.6108 & \mathbf{0}_{1\times 16}\end{pmatrix}$ into (\ref{c2.13}), we obtain $X$ and hence the rest of the observer gains. Accordingly, we obtain the following first-order observer to estimate instantaneously the state variable $x_2(t)$ using measurement vector $\bar{y}_a(t)$
\begin{align}
	\hat{x}_2(t)&=w(t)+\begin{pmatrix}0.2844 &   0.5506   &0\end{pmatrix}\bar{y}_a(t)\nonumber\\&+\begin{pmatrix}-0.0909  & -0.1972 &0\end{pmatrix}\bar{y}_a(t-1)\nonumber\\
	\dot{w}(t)&=w(t)-2.2229w(t-0.5)+w(t-1)\nonumber\\&+\begin{pmatrix}0.8533 &   1.9385  &2 \end{pmatrix}\bar{y}_a(t)\nonumber\\&+\begin{pmatrix}-0.6323 &  -1.2238 &0 \end{pmatrix}\bar{y}_a(t-0.5)\nonumber\\&+\begin{pmatrix}0.0482  & -0.1935  &0 \end{pmatrix}\bar{y}_a(t-1)\nonumber\\&+\begin{pmatrix}0.2022 &   0.4384  &0 \end{pmatrix}\bar{y}_a(t-1.5)\nonumber\\&+\begin{pmatrix}-0.0602  &  0.0909 &0 \end{pmatrix}\bar{y}_a(t-2)+2u(t)\nonumber\\&-1.3856u(t-0.5)+0.4854u(t-1.5).\nonumber
\end{align}

\textit{Case 3:} We look at the case where $0<h<\tau<1\mathrm{s}$. In this situation, we can now employ observer (\ref{c2.17a})-(\ref{c2.17b}) to estimate $z(t)=x_2(t)$. Given that $h<\tau$ and $\tau<1\mathrm{s}$, and as discussed in Case 1 above, we can ensure the asymptotic stability of the error time-delay system
\[
\dot e(t)
=e(t)+N_{\tau}e(t-\tau).
\]

For illustrative purposes, let $h=0.5\mathrm{s}$ and $\tau=0.8\mathrm{s}$. We can now pick, say, $N_{\tau}=-1.03$, which results in the following error time-delay system asymptotically stable
\[
\dot e(t)
=e(t)-1.03e(t-0.8).
\]
Note that by \cite{wu}, the above error time-delay system has two dominant eigenvalues
located at $\{-0.2447\pm j0.1415\}$.

Finally, by substituting $N_{\tau}=-1.03$ into $\Upsilon_r=\begin{pmatrix}FA_{\tau}-N_{\tau}F   &\mathbf{0}\end{pmatrix}$, and solving \eqref{c2.21} for $X_r$, we obtain the following first-order observer to estimate instantaneously the state variable $x_2(t)$
\begin{align}
	\hat{x}_2(t)&=w(t)+\begin{pmatrix}-0.0129 &0.2790\end{pmatrix}y_{\alpha}(t),\nonumber\\
	\dot{w}(t)&=w(t)-1.03w(t-0.8)+\begin{pmatrix}1.9614 &2.0429 \end{pmatrix}y_{\alpha}(t)\nonumber\\&+\begin{pmatrix}-0.5962 &-0.5534 \end{pmatrix}y_{\alpha}(t-0.8)+2u(t)\nonumber\\&-0.545u(t-0.8).\nonumber
\end{align}

For this example, observer $(\ref{c2.17a})$-$(\ref{c2.17b})$ is recommended for the range $0 < h < \tau < 1\text{s}$, as it offers a less complex implementation than $(\ref{c2.5a})$-$(\ref{c2.5b})$.

\section{Conclusion}
The results reported in this paper complement the findings reported in \cite{trinhnam1} for the case where $0<h<\tau$. By employing a novel observer structure together with the augmented measurement vector $y_a(t)$, we have achieved larger allowable delay margins for $\tau$ and $h$.


\begin{thebibliography}{99}
\bibitem{trinhnam1} H. Trinh, P. T. Nam and T. Fernando, ``Existence and design of functional observers for time-delay systems with delayed output measurements'', Preprint at 
https://doi.org/10.48550/arXiv.2603.09395 (2026).

\bibitem{trinhnam2} H. Trinh, P. T. Nam and T. N. Nguyen, ``Time-delay compensators for linear systems with delayed output measurements'', Preprint at https://doi.org/10.48550/arXiv.2604.17434 (2026).

\bibitem{wu} Z. Wu and W. Michiels, ``Reliably computing all characteristic roots of delay differential equations in a given right half plane using a spectral method," \textit{Journal of Computational and Applied Mathematics}, vol. 236, no. 9, pp. 2499-2514, 2012.

\bibitem{mori} T. Mori, ``Criteria for asymptotic stability of linear time-delay systems," \textit{IEEE Transactions on Automatic Control}, vol. 30, no. 2, pp. 158-161, 1985.


\end{thebibliography}
\end{document}